\documentclass[twocolumn,showpacs,preprintnumbers,amsmath,amssymb,aps,prb]{revtex4}

\usepackage{graphicx}
\usepackage{dcolumn}
\usepackage{bm}
\usepackage{color}

\begin{document}


\title{Incoherent Transport through Molecules on Silicon in the vicinity of a Dangling Bond}

\author{Hassan Raza}
 \affiliation{NSF Network for Computational Nanotechnology and School of Electrical and Computer Engineering, Purdue University, West Lafayette, Indiana 47907 USA \\School of Electrical and Computer Engineering, Cornell University, Ithaca New York 14853 USA}%
\author{Kirk H. Bevan, Diego Kienle}
 \affiliation{NSF Network for Computational Nanotechnology and School of Electrical and Computer Engineering, Purdue University, West Lafayette, Indiana 47907 USA}%

\begin{abstract}
We theoretically study the effect of a localized unpaired dangling bond (DB) on occupied molecular orbital conduction through a styrene molecule bonded to a $n^{++}$ H:Si(001)-(2$\times$1) surface. For molecules relatively far from the DB, we find good agreement with the reported experiment using a model that accounts for the electrostatic contribution of the DB, provided we include some dephasing due to low lying phonon modes. However, for molecules within 10$\AA$ to the DB, we have to include electronic contribution as well along with higher dephasing to explain the transport features. 
\end{abstract}


\maketitle

\begin{figure}
\vspace{4in}
\hskip -2.25in\includegraphics{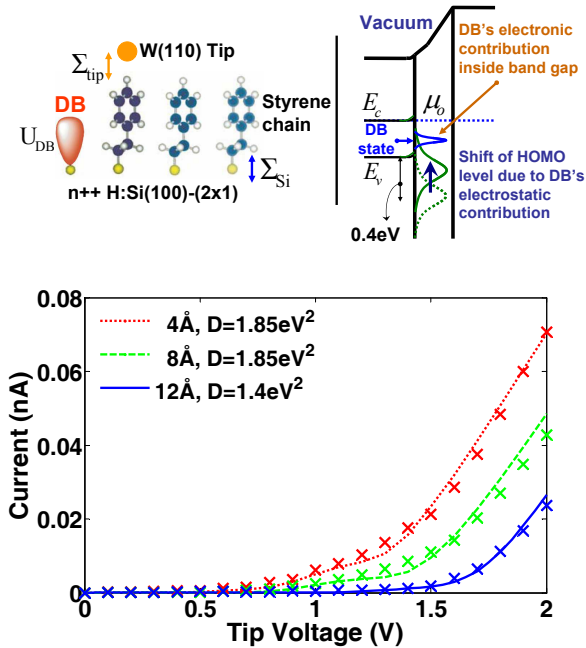}
\caption{(color online) Calculated transport properties of a styrene molecule bonded to a $n^{++}$ H:Si(001)-(2$\times$1) surface in the vicinity of a negatively charged DB. Experimental data is indicated by $\times$ marks \cite{Wolkow05,PaulWolkowDB}. As a function of decreasing distance from the DB, the onset voltages shift toward lower tip voltages implying occupied molecular orbital conduction. I-V plots for molecules 4$\AA$ and 8$\AA$ away have grouped onset voltages. The electrostatic and electronic contributions of the DB are included in these calculations. We assume a lower dephasing strength (D) of 1.4$eV^2$ for the molecule that is farthest from the DB, as compared to 1.85$eV^2$ for the closer ones.}
\end{figure}

\section{Introduction} Interest in atomic scale conduction continues to grow at a rapid pace both in the engineering and scientific communities. The comparatively stable C-Si bond and the well characterized Si surface chemistry \cite{Hamers86,Wolkow88} not only place organic molecules in an ideal position to complement existing Si technology, but also offer a unique platform to study scientific phenomena \cite{Kirczenow05,Liang05} with experimental reproducibility. Notably, Si contacts may give rise to novel features such as negative differential resistance \cite{Titash04,Guisinger05,Bernholc05}, although the mechanism is not yet established \cite{Wolkow06}. 

In Si technology, dangling bond (DB) defects in the form of $P_b$ centers have created reliability issues - consider, negative bias temperature instability in metal-oxide-Si structures. However, one could engineer such defects to be useful in novel devices. Recently Piva \textit{et al} \cite{Wolkow05} observed significant shifts in the onset voltages along a styrene chain as function of distance from a negatively charged unpaired DB on a $n^{++}$ H:Si(001)-(2$\times$1) surface using scanning tunneling spectroscopy (STS). In this paper, we theoretically study their results. A comparison between calculated and experimentally observed current-voltage (I-V) plots is shown in Fig. 1. The DB, being negatively charged, pushes molecular levels upward. A schematic energy level/band diagram is shown in Fig. 1. The onset voltage depends on the gap between the equilibrium chemical potential ($\mu_o$) and molecular levels. Given that the onset voltage is decreasing in the vicinity of the DB, one concludes that conduction is through occupied orbitals. The trend would have been reversed for unoccupied orbital conduction. 

Furthermore, observed onset voltages for molecules 4$\AA$ and 8$\AA$ \cite{PaulWolkowDB} away from the DB are grouped together; whereas a molecule 12$\AA$ away has a distinctly different onset voltage. Electrostatically, one would expect them to follow an inverse distance relationship. To explain the above discrepancy, we propose that the DB also affects the I-V characteristics of the styrene chain electronically by introducing a near-midgap state in the local density of states (LDOS) of Si atoms 4$\AA$ and 8$\AA$ away. For these molecules, this contribution would lead to an early onset of conduction at approximately 0.5V (rather than 1.1V due to the bulk Si band gap) as shown in Fig. 1. Since the DB wavefunction decays exponentially, this contribution would be nearly absent for a molecule 12$\AA$ away. Apart from this, the effect on transport due to the DB state would be pronounced when the highest occupied molecular orbital (HOMO) level is sufficiently broadened to form a significant overlap with the DB state as shown in the inset of Fig. 2. Moreover, this effect would be diminished if the HOMO level is too far from the DB state. 

The impact of dephasing due to low lying phonons on experimentally observed lineshapes is further explored in this paper. In molecular junctions, the dominant phonon modes are the ones having low energy, which arise due to the combined motion [translational, rotational, etc] of the entire molecule relative to the contacts and have been discussed theoretically in Refs. \cite{Guo05,Ventra05}. Experimentally, some evidence of the presence of these low lying modes is available \cite{Shashidhar04, Kawai04, McEuen00}. Since experimental results are suggestive, a detailed study needs to be done to establish it on a solid footing. These low lying phonon modes are excited due to inelastic electron-phonon scattering. Similarly, the styrene should be able to vibrate, rotate and move with respect to the Si and tip contacts, and hence we expect a similar trend. Since $\hbar\omega\ll k_BT$ at room temperature for these modes, we include their effect through elastic dephasing for simplicity. Furthermore, broadening due to such modes could be large as in Ref. \cite{Lannoo98}. The dephasing strength used in our calculation gives a comparable broadening. 

\section{Theoretical Model and Assumptions} We use the single particle non-equilibrium Green's function formalism in the mean field approximation using a non-orthogonal basis to model quantum transport as in Ref. \cite{Ferdows05}. We extend this work by including elastic dephasing within the self-consistent Born approximation \cite{Mahan87,Datta05}. The time-retarted Green's function is defined as:
\begin{eqnarray}G=[(E+i0^+)S-H_d-U_d-U_{DB}-\Sigma_{Si,tip}-\Sigma_{s}]^{-1}\end{eqnarray}
where $H_d$ is the device Hamiltonian and S is the overlap matrix. $U_d$ is the potential experienced by the molecule and consists of (1) Laplace potential due to applied tip voltage, (2) Band lineup potential due to the Fermi level mismatch of n$^{++}$-Si and tip contacts, (3) Hartree and Image potentials, due to non-equilibrium statistics of the electrons on the molecular region. $U_{DB}$ is the potential due to the negative charge on the unpaired dangling bond, which causes a shift in the molecular levels. The electrostatic and electronic boundary conditions are set by the tip voltage and contact self-energies ($\Sigma_{Si,tip}$) respectively. $\Sigma_{s}$ is the scattering self-energy. In a non-orthogonal basis for the elastic dephasing, the scattering broadening function is given as \cite{Ferdows03}: 
\begin{eqnarray}\Gamma_s(E) = \Sigma^{in}_s + \Sigma^{out}_s \approx D\ SA(E)S\end{eqnarray}
where $\Sigma^{in}_s$ is the scattering inflow function and $\Sigma^{out}_s$ is the scattering outflow function. D is the dephasing strength which is a fourth rank tensor, but in our calculations it is approximated by a scalar and is used as a fitting parameter. Treating D as a scalar would ignore the minute details of dephasing that may be otherwise present when a carrier gets scattered from one electronic state to another electronic state more strongly or weakly. A is the Spectral function defined as $i(G-G^{\dagger})=G^n+G^p$, where $G^n$ and $G^p$ are the electron and hole correlation functions respectively. Using Eq. 2, the scattering self-energy is given as:
\begin{equation}\Sigma_s(E)=\frac{1}{\pi}\int_{-\infty}^{\infty}dy\frac{-\Gamma_s(y)/2}{E-y}-i\frac{\Gamma_s(E)}{2}\end{equation}
It has been shown \cite{Magnus06} that the real part of $\Sigma_s(E)$ [calculated by the Hilbert transform as shown in Eq. 2] is about few meV for different organic molecule contacted by Au(111) electrodes due to slowly varying $\Gamma_s(E)$. We expect it to be small for the molecular transport through Si valence band due to slowly varying density of valence band states. However, this approximation may not be appropriate for a one dimensional contact like carbon nanotube or graphene nanoribbon due to van Hove singularities. Given that this calculation is computationally very expensive and has small effect on the final results, we ignore the real part of $\Sigma_s(E)$. Since $\Sigma_s$, G, $G^n (=-iG^<)$ and $\Sigma^{in}_s(=-i\Sigma^<)$ depend on each other, we solve for the above four quantities self-consistently along with the Hatree self-consistent loop of the applied tip voltage. 

\begin{figure}
\vspace{2.6in}
\hskip -2.0in\includegraphics{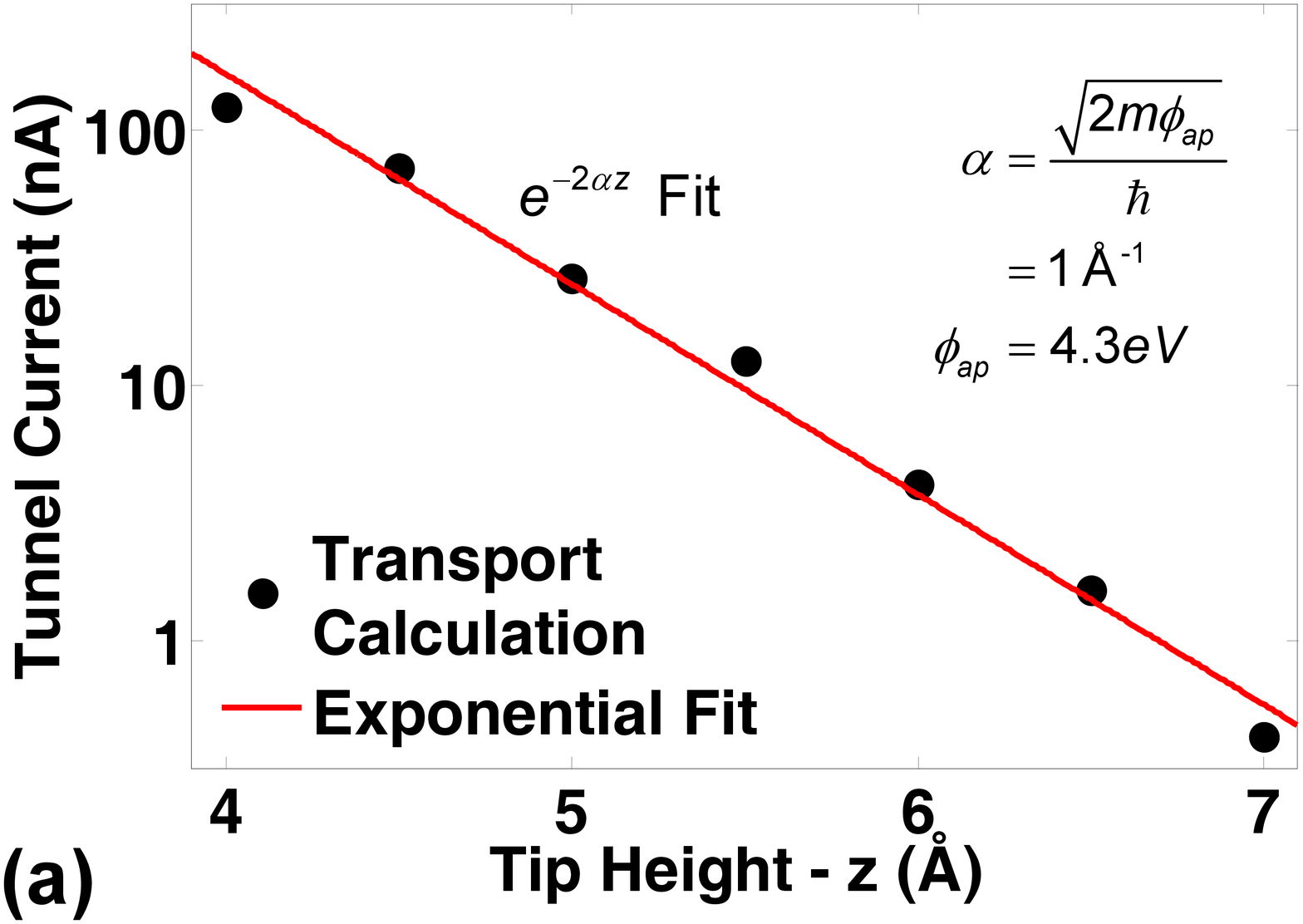}

\vspace{2.6in}
\hskip -2.0in\includegraphics{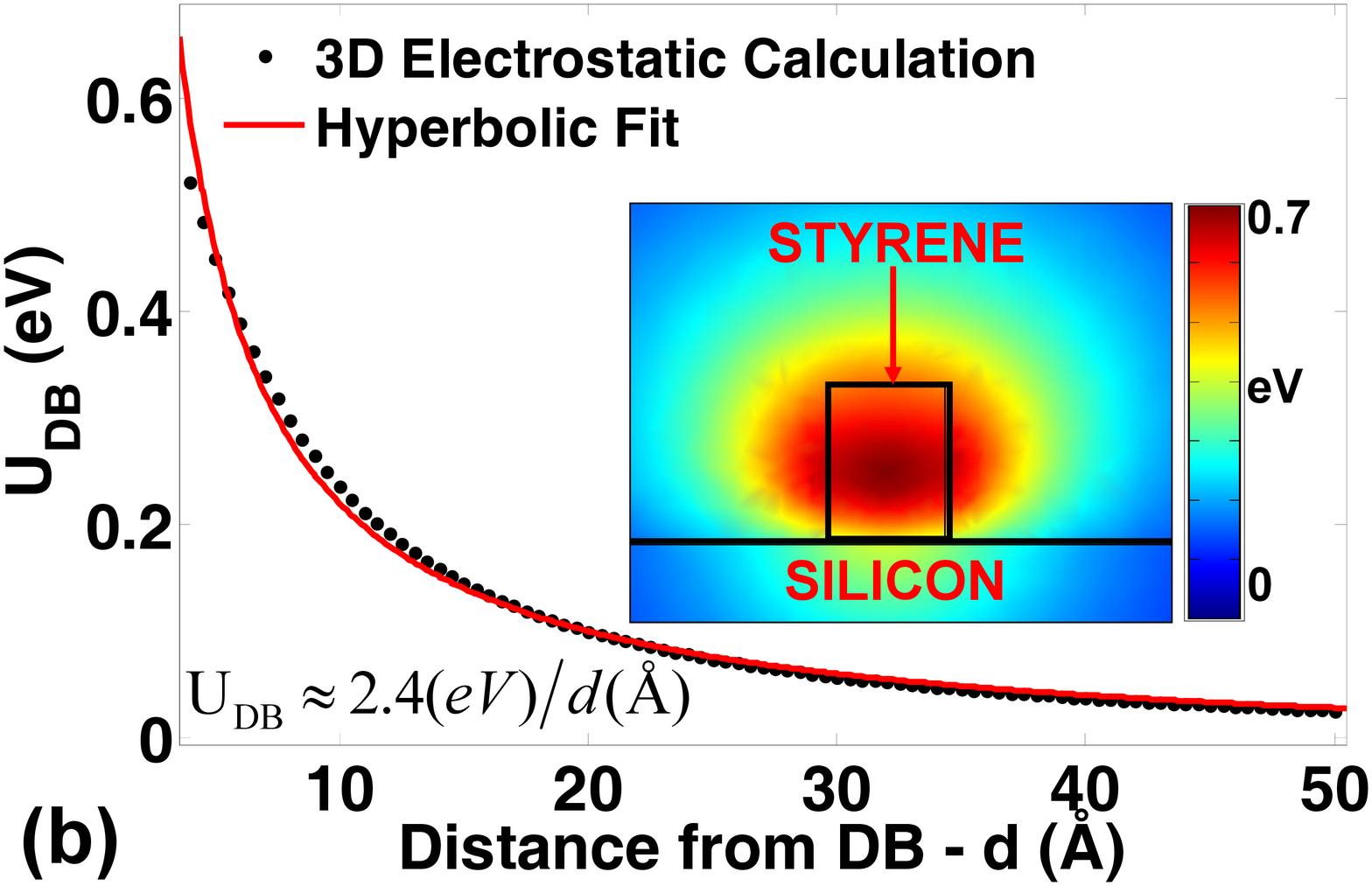}
\caption{(color online) Tip modeling and electrostatic contribution of the DB. (a) Tunnel current as a function of tip height showing exponential decay, which gives an apparent barrier height of 4.3eV. (b) A 3D continuum finite-element electrostatic calculation showing a hyperbolic trend in the average potential, due to a negatively charged DB, along a styrene chain. The insets show a contour plot potential profile 4$\AA$ away.}
\end{figure}

The electronic structure of Si(001) \cite{Cerda00}, styrene \cite{Hoffmann} and the tip is calculated using Extended H\"uckel Theory (EHT). Importantly, EHT provides the correct bulk Si band gap (as compared to LDA \cite{Bernholc05}) and captures the surface and band structure effects accurately \cite{Liang05,Diego06,Raza06}. We incorporate an isolated DB in a H:Si(001)-(2$\times$1) symmetric dimer unit cell, composed of 64 Si atoms with 8 at the surface, out of which 7 are passivated and one is unpassivated which has the dangling bond, as shown in the inset of Fig. 4 (visualized using GaussView \cite{GW03}). Additional surface relaxation due to the DB is small \cite{Watanabe96} and hence ignored. We calculate the surface Green's function ($g_s$) recursively \cite{Rubio84} and find that the DB gives rise to a near-midgap state in the Si band gap similar to Ref. \cite{Watanabe96}. We assume a constant $g_s$ for the tip similar to Ref. \cite{Titash04}. However, $\Sigma_{tip}$ is still energy dependent, since it is defined as $[(E+i0^+)S_1-H_1]g_{s}[(E+i0^+)S_1^\dag-H_1^\dag]$ where $S_1$ and $H_1$ are the overlap and Hamiltonian matrices between tip and styrene molecule. For a good STS tip which can give atomic resolution, the last tip atom at the apex dominates STS and hence we use a single tungsten (W) atom as tip. Transport quantities are then calculated through a single molecule. The molecular structure is taken from Ref. \cite{Titash04}, which reports structure for a styrene molecule on hydrogenated Si surface. There is an important difference between bonding geometry of styrene on hydrogenated and unpassivated Si. For hydrogenated Si, one C atom of styrene bonds to one Si atom as shown in Fig. 1. For unpassivated Si, styrene goes through a cyclo-addition process and the two C atoms on the styrene bind with two atoms of a Si dimer [not shown in this paper]. 

For $\Sigma_{tip}$, the EHT parameters of the s-orbital basis set are modified \cite{TipParameters} (similar to Ref. \cite{Cerda97}) to get an appropriate apparent barrier height ($\phi_{ap}$). We obtain $\phi_{ap}=4.3eV$ from the calculated I(z) plot \cite{Basen96} with 2V applied at tip as shown in Fig. 2(a). This apparent barrier height is about 10eV with original parameters. Apart from this, the HOMO level of W atom lies at -10.73eV, whereas the work function of W(110) tip is around 5.2eV, therefore we have an offset of 5.53eV with respect to vacuum. Similarly, the Si conduction band edge ($E_c$) obtained from EHT has an offset of 7.75eV. To calculate the offset of styrene, we use the HOMO level obtained from DFT-B3PW91/6-311g* calculations \cite{G03} as a reference, since the occupied states in DFT are relatively well characterized \cite{RM04}. To ensure a proper alignment with respect to vacuum, we shift the Hamiltonians by the above-mentioned offsets $E_{offset-i,j}$ for the systems $i$ and $j$ within EHT procedure as follows: 

\begin{equation}H_{ij}=\frac{1}{2}S_{ij}(K_{i}H_{ii}+K_{j}H_{jj}+E_{offset-i}+E_{offset-j})\end{equation}

In order to match the experimentally observed onset voltages, we shift the molecular levels by +0.5eV [shifted toward the valence band minimum]. This shift is a one time adjustment for all the molecules under consideration and hence it affects the onset voltages of all the molecules by an offset of +0.5V. Further changes in the onset voltages come due to the DB potential $U_{DB}$, which varies inversely with distance of different molecules from the DB. Since, the density of states inside HOMO-LUMO (lowest unoccupied molecular orbital) gap is quite small, a shift of 0.5V can be caused by a very small change in Hartree potential at equilibrium due to small charge transfer. However, the lineshapes of the calculated I-V characteristics are not sensitive to this shift and they remain convex [see Sec. III]. Incorporating this shift results in the HOMO level being about 0.4eV below the valence band edge [after incorporating the effect of Si contact as well in the form of Si self-energy - the real part of the self-energy shifts the molecular levels]. This means that HOMO level is about 0.9eV below the valence band edge before applying this shift. The position of HOMO level in our study is different from Ref. \cite{Kirczenow05}, where it is reported to be about 1eV below the valence band edge. However, the difference in our case is that the styrene chain is in the vicinity of a charged dangling bond which could change local nature of Si-C bond. This governs the charge transfer at equilibrium, whereas in Ref. \cite{Kirczenow05}, there is no dangling bond. Furthermore, we do not calculate the charge transfer at equilibrium, rather use the shift in energy levels due to charge transfer as an adjusting parameter. In short, there are two fitting parameters in these calculations, shift of energy levels [a one time adjustment] and dephasing strength D. These two fitting parameters are used to capture two independent features in the experiment. The first one, \textit{i.e.} shift of energy levels, is used to quantitatively match the experimentally observed onset voltages and it does not affect the lineshapes observed whether we include dephasing or not. The second parameter, \textit{i.e.} dephasing strength D, is used to reproduce the experimentally observed \textit{concave} lineshapes. The dephasing affects the onset voltage in a minor fashion since it broadens the energy levels and hence current starts flowing at lower tip voltages. 

\begin{figure}
\vspace{2.5in}
\hskip -2.0in\includegraphics{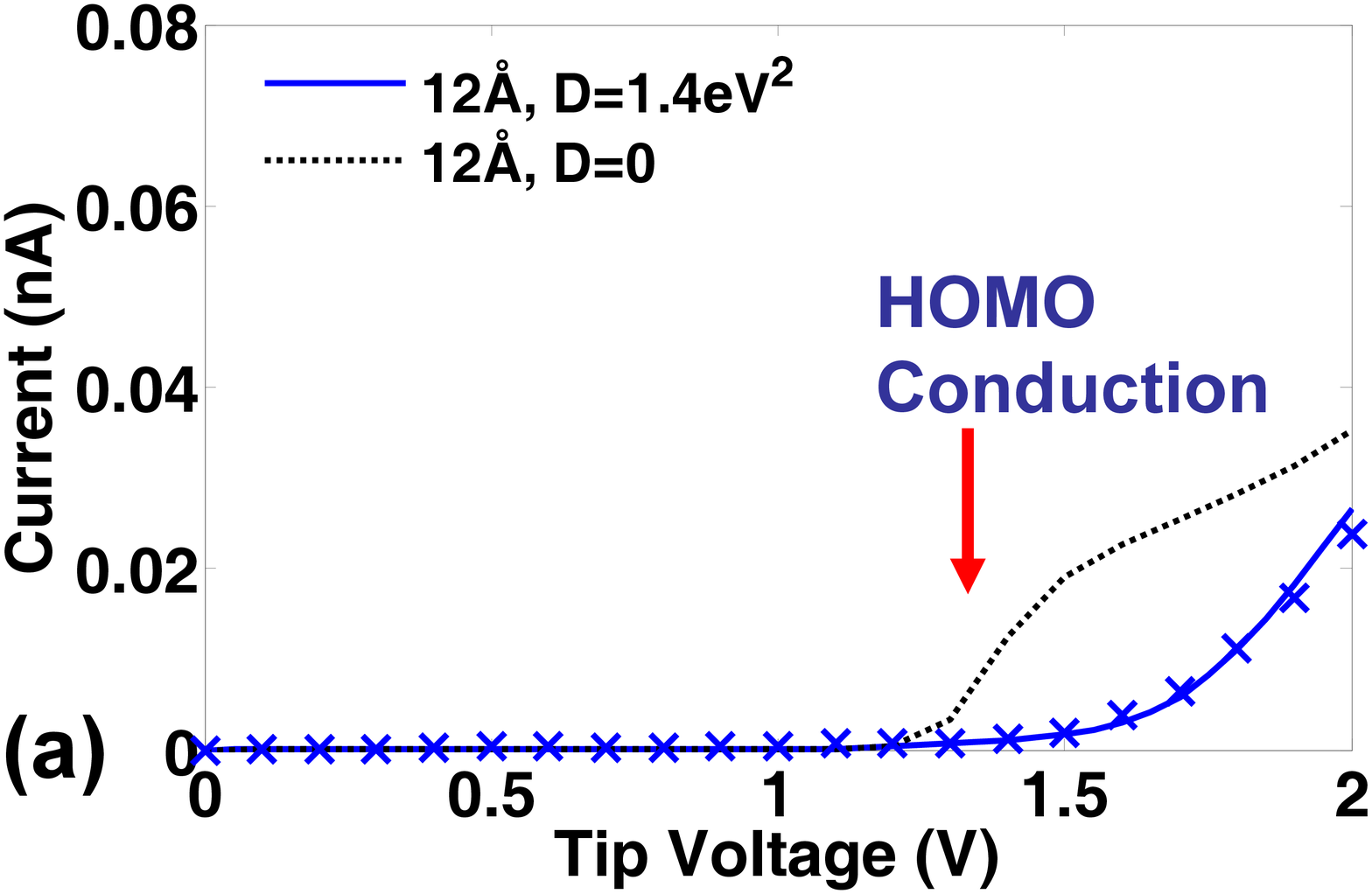}

\vspace{2.5in}
\hskip -2.0in\includegraphics{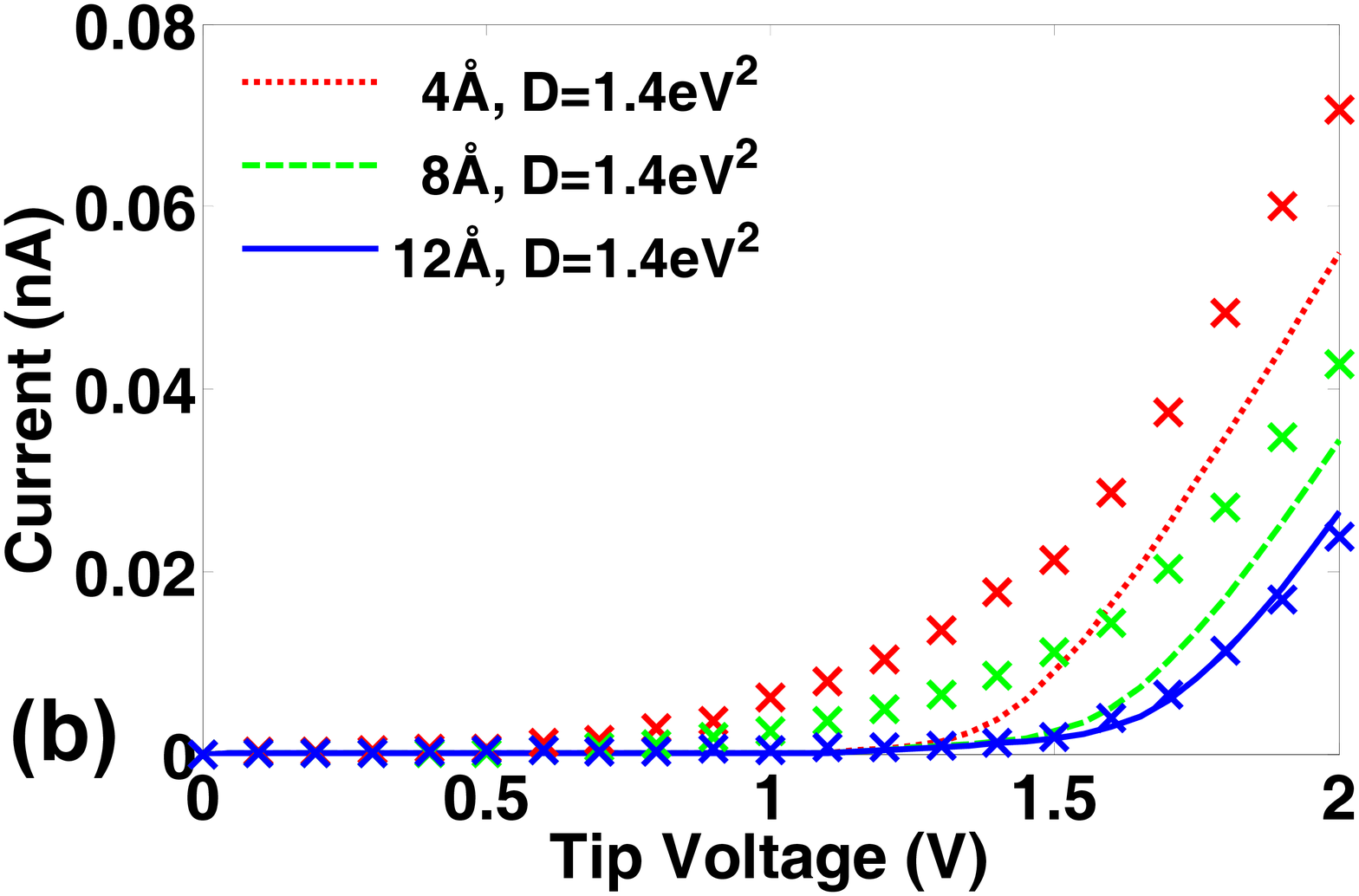}
\caption{(color online) Effect of dephasing and electrostatic contribution of the DB on transport properties. (a) Calculated I-V plots for the molecule 12$\AA$ away with and without dephasing. Without dephasing, the lineshape [red (dark gray) dotted line] does not match with the experiment (blue x marks). With $D=1.4eV^2$, there is a good quantitative agreement with the experiment. (b) Calculated I-V plots for molecules 4$\AA$, 8$\AA$ and 12$\AA$ away from the DB after incorporating $U_{DB}$.}
\end{figure}

We solve the 3D Laplace's equation using the finite-element method to obtain the potential profile across styrene due to the tip voltage. This method has been discussed in detail in Ref. \cite{Ferdows05}. This method also provides the zero bias band lineup potential due to the Fermi level mismatch of 1.15eV between the tip and $n^{++}$ Si. Solving for a 3D potential ensures that Stark effect is captured for the applied tip voltage. Furthermore, for the experimental conditions the molecular levels shift by about one tenth of the applied tip voltage and/or Fermi level mismatch, since the tip is quite far from the styrene molecules. This scenario is different from the ones reported in Refs. \cite{Titash04,Guisinger05}, where the molecular levels shift by about one third of the tip voltage because tip is quite close to the molecule. Apart from this, the tip heights are modified \cite{TipHeight} to reflect actual changes during experimental data acquisition \cite{WolkowPiva} and hence tip heights are not used as adjustment parameters. Rather, it implies that our model is flexible to incorporate the very delicate experimental details.  

For calculating the potential due to the DB, a 3D finite-element continuum calculation \cite{FEMLAB} is performed by taking into account the DB's approximate shape \cite{DB}. The styrene chain is approximated by a 100$\AA$ long 3D box with width and height corresponding to that of styrene and having a relative dielectric constant of about 2.4. This box is placed on a Si bulk having a relative dielectric constant of about 12. An STM tip is placed about 1nm above the styrene chain in the form of a metallic sheet. An average of the calculated potential ($U_{DB}$) taken over the lateral cross-section is included in our transport calculation. A plot of $U_{DB}$ as a function of distance from the DB is shown in Fig. 2(b). The potential profile across a molecule 4$\AA$ away from the DB [shown in the inset of Fig. 2(b)], shows a weak electrostatic contribution in Si. Incorporating $U_{DB}$ in an averaged fashion ignores Stark effect within a single molecule due to charge on the dangling bond. This may quantitatively affect the results in a minor way because as seen in the inset of Fig. 2(b), the dangling bond potential is comparatively constant over a single molecule, \textit{i.e} within about 0.5-0.7eV. This simple continuum model seems to be satisfactory for calculating the electrostatics of the DB, because (1) the screening effect of the styrene chain should be small due to the one dimensional nature of the styrene chain. In particular, occupied states are more localized than unoccupied states and hence this screening is expected to be small for occupied states conduction [a case discussed in this paper] (2) we obtain a hyperbolic dependence as a function of distance from the dangling bond \textit{i.e.} $U_{DB}\approx2.4(eV)/d(\AA)$, which is expected. For calculating the electrostatic contribution, treating the charged DB, styrene chain and $n^{++}$ doped Si in atomistic manner may result in quantitatively different $U_{DB}$, however it is still expected to show a hyperbolic dependence and can be readily incorporated in our model. Such a change would result in different onset voltages as a function of distance from the dangling bond, however the reported lineshapes in Sec. III would remain the same. 

Moreover, based on a finite-element MEDICI \cite{MEDICI} calculation, we conclude that tip induced band bending is also negligible. Additionally, dopants in $n^{++}$ Si introduce states approximately 50meV \cite{Sze} below $E_c$, which contribute weakly \cite{Feenstra87} to transport for a tip voltage of about 50meV. In our calculations, we ignore the above-mentioned effects. The Hartree potential for the molecule is included via the complete neglect of differential overlap method \cite{Pople}. Image effects are incorporated \cite{Ferdows05} to ensure that self-consistency is not over-estimated, which can alter results significantly. In all of the above electrostatic calculations, the tip is taken to be a metal sheet having work function of 5.2eV - a value attributed to W(110) tip. 

\begin{figure}
\vspace{3in}
\hskip -2.0in\includegraphics{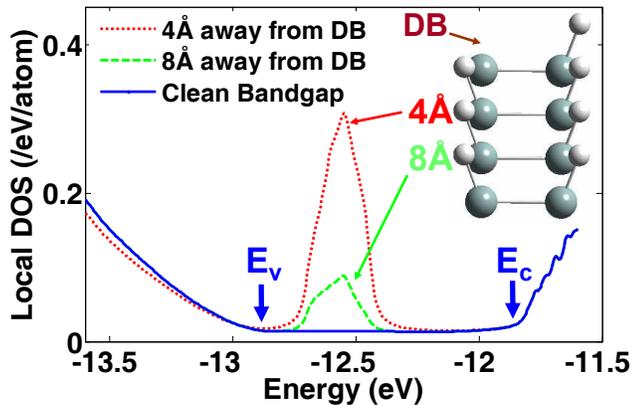}
\caption{(color online) Electronic contribution of the DB. The calculated local density of states of Si atoms, 4$\AA$ and 8$\AA$ from the DB, shows that the DB introduces a near-midgap state. Inset shows the surface portion of the unit cell used.}
\end{figure}

\begin{figure}
\vspace{2.6in}
\hskip -2.0in\includegraphics{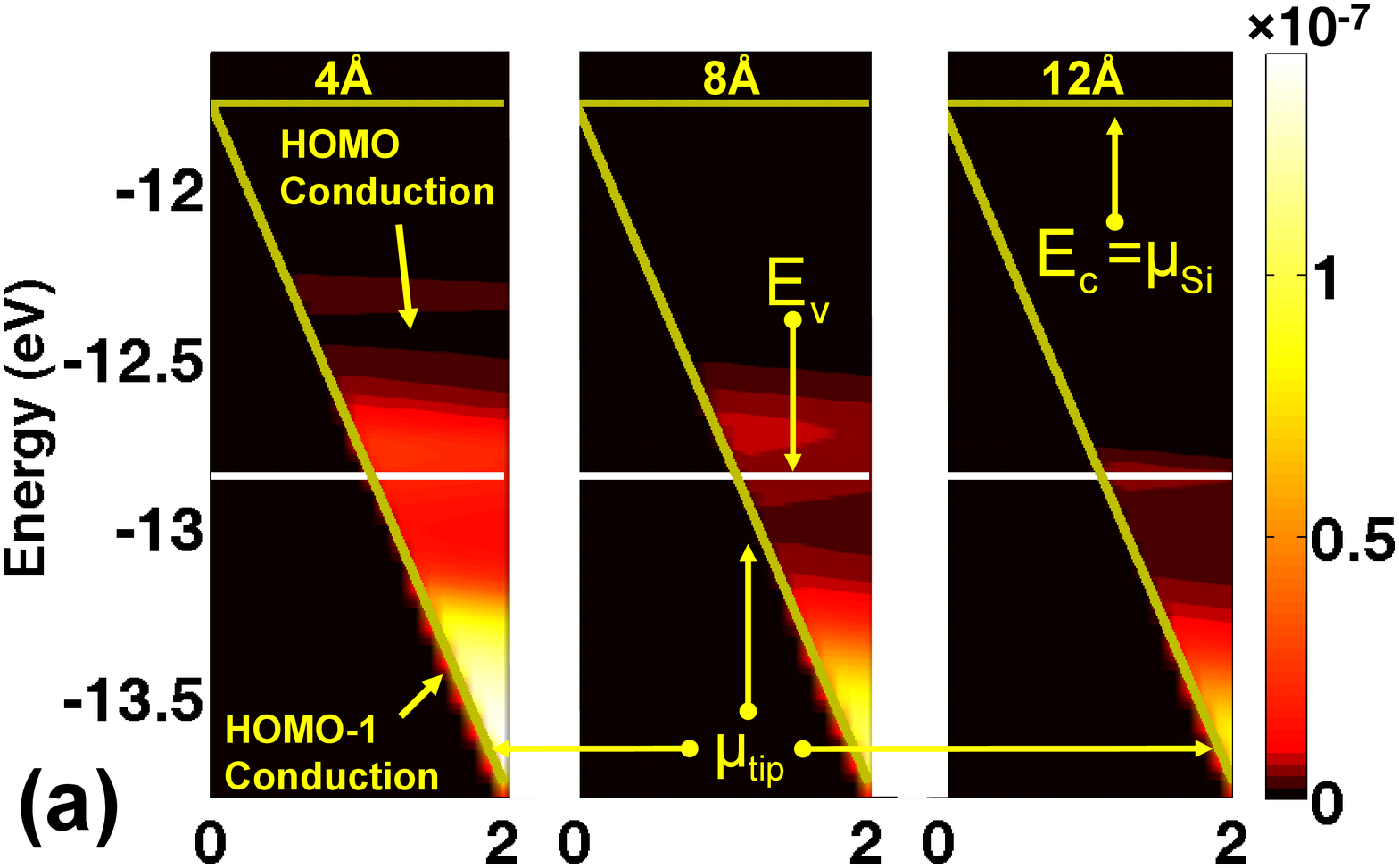}

\vspace{2.5in}
\hskip -2.2in\includegraphics{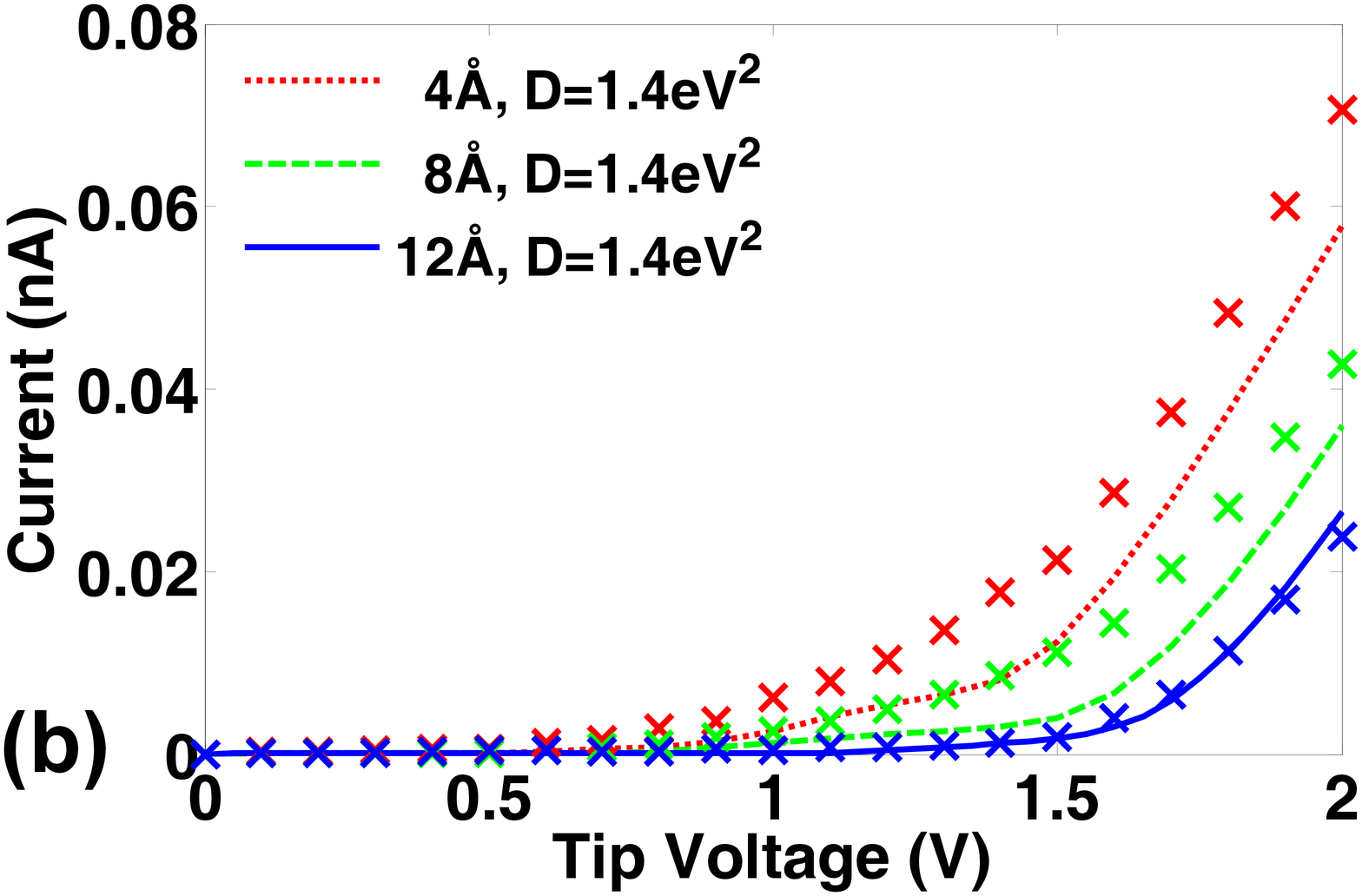}
\caption{(color online) Effect of electronic and electrostatic contributions of the DB on transport properties. (a) $T(E,V)[f_{tip}-f_{Si}]$ plots (where T(E,V) is the transmission and $f_{tip,Si}$ are the Fermi functions for the contacts) for molecules 4$\AA$, 8$\AA$ and 12 $\AA$ away from the DB. For molecules 4$\AA$ and 8$\AA$ away, there is conduction inside band gap due to electronic contribution of the DB. (b) Corresponding I-V plots. Onset voltages for molecules 4 and 8$\AA$ away match well with the experiment, whereas lineshapes do not.}
\end{figure}

\section{Discussion of Results} A transport calculation to study the effect of dephasing for a molecule 12$\AA$ away from the DB is presented in Fig. 3(a). Without dephasing, the lineshape [red (dark gray) dotted line] does not match with the experiment [blue (dark gray) $\times$ marks]. The lineshapes for the theoretically calculated and experimentally observed I-V characteristics are quite different. Experimentally observed lineshapes are \textit{concave}, whereas the calculated lineshape has a \textit{convex} character. For coherent transport, obtaining a \textit{convex} lineshape is expected because it represents that current starts flowing through a level and then saturates when the level is fully inside the bias window. The corresponding conductance peak [not shown] has a full width half maximum (FWHM) of about 0.2eV. This is roughly equal to the coupling between styrene and Si contact because tip is quite far and hence the I-V characteristics do not get affected much by the Hartree potential due to the non-equilibrium carrier statistics. This small coupling is expected since styrene binds with Si through only one Si-C bond and HOMO is localized on the aromatic ring of the styrene molecule, see \textit{e.g.} Refs. \cite{Wolkow05,Wolkow00}. However, after including dephasing (with $D=1.4eV^2$), not only is the lineshape reproduced, but good quantitative agreement with the experiment is achieved. 

This transition from a convex lineshape to concave lineshape needs further discussion. The physical effect of dephasing is that it broadens the molecular levels. Since the local density of states (LDOS) of Si contact [see Fig. 4] increases with decreasing energy/increasing positive tip voltage, the current starts increasing in a fashion proportional to the coupling to the Si contact, which is proportional to the LDOS of Si contact. In this case, as can be seen in Fig. 4, the LDOS increases with decreasing energy as one goes away from the valence band edge. The current increases with increasing tip voltage in a concave manner because of this increase in LDOS of Si contact. This LDOS corresponds to increased surface Green's function of Si and hence larger Si self-energy $\Sigma_{Si}$, which results in a higher current. In order to adjust for this, we have to use higher tip height of 8.67$\AA$ in our model, after including dephasing, instead of 7.79$\AA$ for coherent transport to get the same current level. Since experimentally tip heights in general remain to be unknown, this variation is acceptable. However, relative tip heights are known experimentally. In the set of experiments we consider, these relative heights were changed \cite{WolkowPiva} during experimental data acquisition for the molecules 4$\AA$ and 8$\AA$ away from the DB and we include them in our model \cite{TipHeight}. 

Using the same D for molecules 4$\AA$ and 8$\AA$ away from the DB, and including $U_{DB}$, we obtain the I-V plots shown in Fig. 3(b). Although the lineshapes are similar to the experimentally observed lineshapes, there is still a significant mismatch in the onset voltages between experiment and calculation for molecules 4$\AA$ and 8$\AA$ away. Experimentally, their onset voltages are grouped together and are much smaller than the calculated ones. This is expected since the valence band edge corresponds to a tip voltage of about 1.1V and hence very small current should be expected for tip voltages below 1.1V. However, as can be seen in Fig. 3(b), experimentally conduction starts from about 0.5V. We interpret this disagreement between experiment and calculation to be induced by the DB. Besides contributing electrostatically, the DB introduces a near-midgap state in the LDOS of Si atoms beneath the styrene chain extending up to be about 10$\AA$. The calculated LDOS for Si atoms 4$\AA$ and 8$\AA$ away from the DB is shown in Fig. 4. A near-midgap state is clearly visible inside the band gap which is responsible for the early onset of conduction at about 0.5V of tip voltage. The local DOS contribution of the DB state exponentially decays and is nearly absent for a styrene molecule about 12$\AA$ away. 

In our transport model, the DB is taken to be fully occupied under equilibrium and non-equilibrium conditions through inelastic processes inside the Si contact. This is a valid assumption since the energetic location of the DB state is well below the chemical potential. Furthermore, the electronic structure calculations are for intrinsic Si, where the effect of dopants and charged DB is ignored for simplicity, otherwise the calculations become computationally prohibitive. The effect of doping, however, is electrostatically included in our transport calculations by adjusting the equilibrium Si chemical potential. Besides this, treating the DB state as charged is expected to cause a shift in its energetic position within band gap. Since these states are thermally broadened on the order of 0.5eV or more \cite{Hitosugi} at room temperature and that the DB state is close to midgap, the results presented are expected to be comparatively insensitive to this shift. 

In Fig. 5(a), the bias dependent calculated transmission T(E,V) through molecules 4$\AA$, 8$\AA$ and 12$\AA$ away from the DB is shown. For molecules 4$\AA$ and 8$\AA$ away, there is finite conduction inside the band gap due to this DB state, which is absent for the molecule 12$\AA$ away. For the molecule 4$\AA$ away, at higher tip voltages, the level below the HOMO level \textit{i.e.} HOMO-1 level, also starts conducting. Fig. 5(b) shows the corresponding I-V characteristics after including the DB's electronic and electrostatic contribution with $D=1.4eV^2$. The onset voltages for molecules 4$\AA$ and 8$\AA$ away match reasonably well with the experiment and are grouped together, but the lineshapes still differ. There is an increase in current at about 0.6V and then current tends to saturate. This is because contribution of the DB state is small in this energy range. Using a higher D of 1.85$eV^2$ for molecules 4$\AA$ and 8$\AA$ away broadens the molecular spectral density more and hence reproduces the experimentally observed lineshapes as shown in Fig. 1. These molecules, being at the end of styrene chain, may have more phonon degrees of freedom thus resulting in a higher dephasing. Another possible explanation could be that the dephasing processes inside the Si contact broaden the LDOS contribution of the midgap state. Further experimental and theoretical work needs to be carried out to establish a detailed understanding of the dephasing mechanism. For example, conducting temperature dependent transport experiments would sharpen the spectroscopic features with decreasing temperature, since $D\propto T$ for $\hbar\omega\ll k_BT$. Apart form this, using different molecules with varying HOMO levels may give an insight into the electronic contribution.

Since we are calculating transport properties of an isolated styrene molecule and comparing with the styrene molecules embedded in a chain, the question arises if we are missing something. As far as one can obtain atomic resolution with the tip which means that there is one apex atom and the tip is above a styrene molecule, the conduction from the tip to other styrene molecules inside the chain is expected to be small. Because styrene molecules are about 4$\AA$ away and if the tip is about 8$\AA$ above the styrene molecule, the distance of the tip atom from the next styrene molecule would be about 9$\AA$. This should result in an order of magnitude lower current through the next styrene molecule. Therefore, although the conduction along the styrene chain may be noticeable as discussed in Ref. \cite{Kirczenow05}, since the tip is placed above a styrene molecule, the lowest conduction path is still through this particular styrene molecule and hence conduction through the neighboring styrene molecules would still be small. However, if the tip is placed between two styrene molecule [a case not discussed in this paper], one has to consider transport though both the styrene molecules as well as tunneling between them. Since in our case, the neighboring styrene molecules are expected to remain in equilibrium, the effect of self-consistent potential variation in the neighboring styrene molecules can be ignored as well. It should be noted that our case is different from that of a self-assembled monolayer device contacted by a flat contact, because in this case all the molecules are expected to conduct and hence a molecule would be affected by the self-consistent potential of the neighboring molecules. The self-consistent potential of the styrene molecule through which the conduction is occurring could affect the neighboring styrene molecule electrostatics and in return the perturbed neighboring molecules could affect the self-consistent potential of the conducting molecule itself. Since, tip is far from the molecule and self-consistent potential is small [although not negligible], this effect is expected to be small. 

We also ignore the tilt angle of the styrene molecule with respect to the Si substrate as a function of distance from the dangling bond that may be present in reality. It has been shown previously that the tilt angle for a styrene chain in the absence of the dangling bond as a function of distance from the end of the styrene chain may lead to significant changes in the current flowing through the styrene chain and hence the observed STM images. This end effect has been observed experimentally for low bias and qualitatively reproduced theoretically in Ref. \cite{Kirczenow05}. This end effect vanishes at higher bias and is still not well understood. As shown experimentally in Ref. \cite{Wolkow05}, in the presence of a charged dangling bond at the end of the styrene chain, this end effect is not present. Therefore, assuming that the tilt angle of the molecules near the end of the chain does not change seem to be a good approximation. However, a detailed quantitative study is needed. It would also be interesting to analyze how the low lying phonon spectrum gets modified for a styrene chain on Si surface in the presence of a dangling bond, whereas previously phonon spectrum of a single molecule on a substrate has been discussed \cite{Guo05,Ventra05}. The inter-molecular interactions are expected to give rise to new peaks in the phonon spectrum. 

Apart from this, there is a possibility that the neighboring styrene molecules may be electronically affecting each other. Although the styrene molecules are not chemically bonded, there still may be some effect of broadening. In particular, this effect has been theoretically addressed in Ref. \cite{Kirczenow05} where they conclude that it would be higher for unoccupied states than occupied states. This conclusion is consistent with Ref. \cite{Wolkow02}, where based on DFT calculations, it has been reported that for occupied states upto 80\% of the wave-function is localized on the aromatic ring of the styrene molecule. Furthermore, the unoccupied states are also analyzed theoretically in Refs. \cite{Wolkow00,Cho02} reaching the same conclusion. It means that hybridization along the chain for occupied state is small and hence broadening caused by it is expected to be small. Another calculation in Ref. \cite{Wolkow05} shows similar behavior with additional information that for the energy range correponding to tip voltages less than about +2V [the voltage range discussed in this paper], the coupling between the styrene molecules is small and hence they attribute the experimental observation of the slope effect to this localized charge on styrene molecules in this voltage range. However, for larger energy range and for tip voltages greater than about +2V [a case not discussed in this paper], the spectral densities on neighboring styrene molecules start overlaping and charge density becomes delocalized and hence slope effect vanishes. Since in our study, we are calculating transport properties for occupied states in 0-2V tip voltage range, this effect is expected to be small in our voltage range of interest and hence is not included in our calculations.

\section{Conclusions} We have studied occupied level conduction through a styrene molecule in the vicinity of a negatively charged DB on a $n^{++}$ H:Si(001)-(2$\times$1) surface. We put forward that the DB not only affect conduction through styrene electrostatically within approximately $100\AA$ but also electronically up to $10\AA$ from the DB by introducing a near-midgap state in the LDOS of neighboring Si atoms. Dephasing is further expected to play a significant role in these experiments. 

\section{Appendix}

In this section we show the derivation of Eq. 2. The scattering inflow function ($\Sigma^{in}_s$), outflow function ($\Sigma^{out}_s$) and broadening function ($\Gamma_s$) are defined as \cite{Mahan87}:

\begin{equation} 
\Sigma^{in,out}_s(E)=\int_0^{\infty}\frac{d(\hbar\omega)}{2\pi}D^{em,ab}(\hbar\omega)SG^{n,p}(E+\hbar\omega)S+D^{ab,em}(\hbar\omega)SG^{n,p}(E-\hbar\omega)S\nonumber\end{equation}
\begin{equation}\Gamma_s(E)=\Sigma^{in}_s+\Sigma^{out}_s\nonumber\end{equation}
\begin{equation}=\int_0^{\infty}\frac{d(\hbar\omega)}{2\pi}D^{em}(\hbar\omega)S[G^n(E+\hbar\omega)+G^p(E-\hbar\omega)]S+D^{ab}(\hbar\omega)S[G^n(E-\hbar\omega)+G^p(E+\hbar\omega)]S\nonumber\end{equation}
where $D^{em}(\hbar\omega)=(N+1)D_o(\hbar\omega)$ and $D^{ab}(\hbar\omega)=N\ D_o(\hbar\omega)$ are emission and absorption dephasing functions respectively and are related by $D^{ab}/D^{em}=exp(-\hbar\omega / k_BT)$. N is the equilibrium number of phonons given by Bose-Einstein statistics as $1/[exp(\hbar\omega / k_BT)-1]$. For $\hbar\omega\ll k_BT$, $N+1 \approx N \approx {k_BT}/{\hbar\omega}$ by keeping first term in the Taylor series expansion of $exp(\hbar\omega / k_BT)\approx 1+\hbar\omega / k_BT$, which leads to,
\begin{eqnarray}\Gamma_s(E)=\int_0^{\infty}\frac{d(\hbar\omega)}{2\pi}[D^{em}(\hbar\omega)+D^{ab}(\hbar\omega)]SA(E)S \nonumber\end{eqnarray}
and 
\begin{eqnarray}D^{em}(\hbar\omega)\approx D^{ab}(\hbar\omega)=N\ D_o(\hbar\omega)\approx \frac{k_BT}{\hbar\omega} D_o(\hbar\omega)\nonumber\end{eqnarray}
Substituting these results in the above Eq. gives Eq. 2:
\begin{eqnarray}\Gamma_s(E) \approx \underbrace{T\int_0^{\infty}\frac{d(\hbar\omega)}{2\pi}\frac{2k_BD_o(\hbar\omega)}{\hbar\omega}}_{D}SA(E)S\nonumber\end{eqnarray}
The presence of low lying phonon modes depends on many details (type of molecule, surface, bias voltage, etc). This implies that not only the energy ($\hbar\omega$) of the phonon modes would change under different conditions, but also the associated dephasing functions $D_o$. We leave the detailed study of these phonon modes for the styrene chain on Si surface in the presence of a dangling bond for future work and in this study, treat this effect in an average manner by using the dephasing strength D as an adjusting parameter. 

We acknowledge fruitful discussions with S. Datta, P. G. Piva and R. A. Wolkow and thank the latter two for sharing their experimental data in electronic format. We acknowledge F. Zahid and T. Raza for Huckel-IV 3.0 \cite{Ferdows05} codes. We also thank A. W. Ghosh and G.-C. Liang for useful discussions. This work was supported by the NASA Institute for Nanoelectronics and Computing and ARO-DURINT. Computational facilities were provided by the NSF Network for Computational Nanotechnology.


\begin{thebibliography}{100}
\bibitem{Hamers86} R. J. Hamers, R. M. Tromp and J. E. Demuth, Phys. Rev. B {\bf{34}}, 5343 (1986).

\bibitem{Wolkow88} R. Wolkow and Ph. Avouris, Phys. Rev. Lett. {\bf{60}}, 1049 (1988).

\bibitem{Kirczenow05} G. Kirczenow, P. G. Piva and R. A. Wolkow, Phys. Rev. B {\bf{72}}, 245306 (2005). 
\bibitem{Liang05} G.-C. Liang and A. W. Ghosh, Phys. Rev. Lett. {\bf{95}}, 076403 (2005) and references there-in.
\bibitem{Titash04} T. Rakshit, G.-C. Liang, A. W. Ghosh and S. Datta, Nano. Lett. {\bf{4}}, 1803 (2004). 
\bibitem{Guisinger05} N. P. Guisinger, N. L. Yolder and M. C. Hersam, PNAS {\bf{102}}, 8828 (2005). 
\bibitem{Bernholc05} W. Lu, V. Meunier and J. Bernholc, Phys. Rev. Lett. {\bf{95}}, 206805 (2005).
\bibitem{Wolkow06} J. L. Pitters and R. A. Wolkow, Nano. Lett. {\bf{6}}, 390 (2006).
\bibitem{Wolkow05} P. G. Piva, G. A. DiLabio, J. L. Pitters, J. Zikovsky, M. Rezeq, S. Dogel, W. A. Hofer and R. A. Wolkow, Nature {\bf{435}}, 658 (2005).
\bibitem{PaulWolkowDB} P. G. Piva and R. A. Wolkow (private communication). Experimentally, there is an uncertainty of $\pm$ half a dimer spacing to determine the DB centre. Therefore, $4\AA$ and $8\AA$ curves are more like $4\pm2\AA$ and $8\pm2\AA$.

\bibitem{Guo05} N. Sergueev, D. Roubtsov and H. Guo, Phys. Rev. Lett. {\bf{95}}, 146803 (2005).
\bibitem{Ventra05} Y.-C. Chen, M. Zwolak and M. D. Ventra, Nano. Lett. {\bf{5}}, 621 (2005).

\bibitem{Shashidhar04} J. G. Kushmerick, J. Lazorcik, C. H. Patterson, R. Shashidhar, D. S. Seferosand G. C. Bazan, Nano. Lett. {\bf{4}}, 639 (2004).

\bibitem{Kawai04} H. S. Kato, J. Noh, M. Hara and M. Kawai, J. Phys. Chem. B {\bf{106}}, 9655 (2002).

\bibitem{McEuen00} H. Park, J. Park, A. K. L. Lim, E. H. Anderson, A. P. Alivisatos, P. L. McEuen, Nature {\bf{407}}, 57 (2000).

\bibitem{Lannoo98} A. Devos and M. Lannoo, Phys. Rev. B {\bf{58}}, 8236 (1998).
\bibitem{Ferdows05} F. Zahid, M. Paulsson, E. Polizzi, A. W. Ghosh, L. Siddiqui, and S. Datta, J. Chem. Phys. {\bf{123}}, 064707 (2005).
\bibitem{Datta05} S. Datta, \textit{Quantum Transport: Atom to Transistor} (Cambridge University Press, Cambridge, UK, 2005).
\bibitem{Mahan87} G. D. Mahan, Phys. Rep. {\bf{145}}, 251 (1987).
\bibitem{Ferdows03} F. Zahid, M. Paulsson and S. Datta, "Electrical Conduction through Molecules", in \textit{Advanced Semiconductors and Organic Nano-techniques (III)}, edited by H. Morkoc (Academic Press, San Diego, CA, 2003).
\bibitem{Magnus06} M. Paulsson, T. Frederiksen and M. Brandbyge, Nano Lett. {\bf{6}}, 258 (2006).
\bibitem{Cerda00} J. Cerda and F. Soria, Phys. Rev. B {\bf{61}}, 7965 (2000).
\bibitem{Hoffmann} J. Howell, A. Rossi, D. Wallace, K. Haraki and R. Hoffman, FORTICON8, QCPE Program 545, Department of Chemistry, Cornell University, Ithaca, NY 14853.
\bibitem{Raza06} H. Raza, Phys. Rev. B {\bf{76}}, 045308 (2007).
\bibitem{Diego06} D. Kienle, K. H. Bevan, G.-C. Liang, L. Siddiqui, J. I. Cerda and A. W. Ghosh, J. Appl. Phys. {\bf{100}}, 043715 (2006).
\bibitem{Watanabe96} S. Watanabe, Y. A. Ono, T. Hashizume and Y. Wada, Phys. Rev. B {\bf{54}}, R17308 (1996). 
\bibitem{GW03} R. Dennington II, T. Keith, J. Millam, K. Eppinnett, W. L. Hovell, and R. Gilliland, \textit{GaussView, Version 3.0} (Semichem, Inc., Shawnee Mission, KS, 2003).
\bibitem{Rubio84} M. P. L. Sancho, J. M. L. Sancho and J. Rubio, J. Phys. F: Met. Phys. {\bf{14}}, 1205 (1984).
\bibitem{TipParameters} Original $\zeta$=2.341, Modified $\zeta$=1.10427.
\bibitem{Cerda97} J. Cerda, A. Yoon, M. A. VanHove, P. Sautet, M. Salmeron and G. A. Somorjai, Phys. Rev. B {\bf{56}}, 15900 (1997).
\bibitem{Basen96} L. Olesen, M. Brandbyge, M. R. Sorensen, K. W. Jacobsen, E. Laegsgaard, I. Stensgaard and F. Besenbacher, Phys. Rev. Lett. {\bf{76}}, 1485 (1996).
\bibitem{G03} M. J. Frisch, G. W. Trucks, H. B. Schlegel, G. E. Scuseria, M. A. Robb, J. R. Cheeseman, J. A. Montgomery, Jr., T. Vreven, K. N. Kudin, J. C. Burant, J. M. Millam, S. S. Iyengar, J. Tomasi, V. Barone, B. Mennucci, M. Cossi, G. Scalmani, N. Rega, G. A. Petersson, H. Nakatsuji, M. Hada, M. Ehara, K. Toyota, R. Fukuda, J. Hasegawa, M. Ishida, T. Nakajima, Y. Honda, O. Kitao, H. Nakai, M. Klene, X. Li, J. E. Knox, H. P. Hratchian, J. B. Cross, V. Bakken, C. Adamo, J. Jaramillo, R. Gomperts, R. E. Stratmann, O. Yazyev, A. J. Austin, R. Cammi, C. Pomelli, J. W. Ochterski, P. Y. Ayala, K. Morokuma, G. A. Voth, P. Salvador, J. J. Dannenberg, V. G. Zakrzewski, S. Dapprich, A. D. Daniels, M. C. Strain, O. Farkas, D. K. Malick, A. D. Rabuck, K. Raghavachari, J. B. Foresman, J. V. Ortiz, Q. Cui, A. G. Baboul, S. Clifford, J. Cioslowski, B. B. Stefanov, G. Liu, A. Liashenko, P. Piskorz, I. Komaromi, R. L. Martin, D. J. Fox, T. Keith, M. A. Al-Laham, C. Y. Peng, A. Nanayakkara, M. Challacombe, P. M. W. Gill, B. Johnson, W. Chen, M. W. Wong, C. Gonzalez, and J. A. Pople, \textit{Gaussian 03, Revision B.03} (Gaussian, Inc., Wallingford CT, 2004).
\bibitem{RM04} R. M. Martin, \textit{Electronic Structure: Basic Theory and Practical Methods} (Cambridge University Press, Cambridge, UK, 2004).
\bibitem{TipHeight} 8.79$\AA$, 8.73$\AA$ and 8.67$\AA$ for molecules 4$\AA$, 8$\AA$ and 12$\AA$ away respectively unless otherwise stated.
\bibitem{WolkowPiva} P. G. Piva and R. A. Wolkow (private communication). During the constant current STM (not shown), the height profile of the styrene chain decreased $1.2\pm0.4\AA$ across its 60$\AA$ length. As this constant current height contour (with the current feedback loop 'ON') determined the starting tip-sample separation prior to acquiring individual I-V curves (with current feedback loop 'OFF'), I-V curves shown in Fig. 1, 3, and 5 in the present work and in Fig. 5(a) in Ref. \cite{Wolkow05} were acquired with decreasing tip-sample separation as the distance from the DB increased. 
\bibitem{FEMLAB} COMSOL is a trademark of COMSOL AB (comsol.com).
\bibitem{DB} DB is approximated by a prolate spheroid with polar and equatorial radius given by 1.5$\AA$ and 1$\AA$ respectively, estimated with a B3PW91/6-311g* calculation. The relative dielectric constant for the dangling bond is taken as 1. 
\bibitem{MEDICI} Technology Modeling Associates, Inc.: Sunnyvale, CA, \textit{TMA medici, two-dimensional device simulation program, version 4.0 user's manual} (1997). 
\bibitem{Sze} S. M. Sze, \textit{Physics of Semiconductor Devices Ch. 1} (Wiley-Interscience, NY, 1981).
\bibitem{Feenstra87} R. M. Feenstra and J. A. Stroscio, J. Vac. Sci. Technol. B {\bf{5}}, 923 (1987).
\bibitem{Pople} J. A. Pople and G. A. Segal, J. Chem. Phys. {\bf{44}}, 3289 (1966).
\bibitem{Hitosugi} T. Hitosugi, T. Hashizume, S. Heike, Y. Wada, S. Watanabe, T. Hasegawa, K. Kitazawa, Appl. Phys. A {\bf{66}}, S695 (1998).
\bibitem{Wolkow02} W. A. Hofer, A. J. Fisher, G. P. Lopinski and R. A. Wolkow, Chem. Phys. Lett. {\bf{365}}, 129 (2002).

\bibitem{Wolkow00} G. P. Lopinski, D. D. M. Wayner and R. A. Wolkow, Nature {\bf{406}}, 48 (2000).

\bibitem{Cho02} J.-H. Cho, D.-H. Oh and L. Kleinman, Phys. Rev. B {\bf{65}}, 081310(R) (2002).

\end{thebibliography}
\end{document}